\documentclass[11pt]{article}
\usepackage[utf8]{inputenc}
\usepackage{amsfonts}
\usepackage{amsmath}
\usepackage{amssymb}
\usepackage{indentfirst}
\usepackage{graphicx}
\usepackage[colorlinks]{hyperref}
\usepackage{cite}

\usepackage{microtype}
\usepackage[normalem]{ulem}

\numberwithin{equation}{section}
\allowdisplaybreaks

\begin{document}

\begin{titlepage}
\vspace{3cm}

\baselineskip=24pt

\begin{center}
\textbf{\LARGE{Three-dimensional teleparallel Chern-Simons supergravity theory}}
\par\end{center}{\LARGE \par}

\begin{center}
	\vspace{1cm}
	\textbf{Ricardo Caroca}$^{\ast}$,
	\textbf{Patrick Concha}$^{\ast}$,
    \textbf{Diego Peñafiel}$^{\ddag}$,
	\textbf{Evelyn Rodríguez},
	\small
	\\[5mm]
    $^{\ast}$\textit{Departamento de Matemática y Física Aplicadas, }\\
	\textit{ Universidad Católica de la Santísima Concepción, }\\
\textit{ Alonso de Ribera 2850, Concepción, Chile.}
	\\[2mm]
	$^{\ddag}$\textit{Facultad de Ciencias, Universidad Arturo Prat, }\\
	\textit{Iquique, Chile.}
	\\[5mm]
	\footnotesize
	\texttt{rcaroca@ucsc.cl},
	\texttt{patrick.concha@ucsc.cl},
    \texttt{dimolina@unap.cl},
	\texttt{everodriguez@gmail.com},
	\par\end{center}
\vskip 26pt
\begin{abstract}
In this work we present a gauge-invariant three-dimensional teleparallel supergravity theory using the Chern-Simons formalism. The present construction is based on a supersymmetric extension of a particular deformation of the Poincaré algebra. At the bosonic level the theory describes a non-Riemannian geometry with a non-vanishing torsion. In presence of supersymmetry, the teleparallel supergravity theory is characterized by a non-vanishing super-torsion in which the cosmological constant can be seen as a source for the torsion. We show that the teleparallel supergravity theory presented here reproduces the Poincaré supergravity in the vanishing cosmological limit. The extension of our results to $\mathcal{N}=p+q$ supersymmetries is also explored. 
\end{abstract}
\end{titlepage}\newpage {} {\baselineskip=12pt }

\section{Introduction}

Teleparallel gravity is an alternative theory of gravity known to be considered equivalent to General Relativity. However, they are conceptually quite different. In particular, the teleparallel formulation of gravity is described by a vanishing curvature and a non-vanishing torsion which characterizes the parallel transport \cite{Hayashi:1979qx,Kawai:1993kvr,deAndrade:1997gka,Sousa:2000bc,DeAndrade:2000sf}. In such case, the geometry is no more Riemannian but corresponds to the so-called Riemannian-Cartan (Weizenböck) geometry.

In three spacetime dimensions, there has been an interest in exploring black hole solutions and boundary symmetries of gravity theories with torsion \cite{Garcia:2003nm,Mielke:2003xx,Blagojevic:2003vn,Blagojevic:2003uc,Blagojevic:2003wn,Blagojevic:2013bu,Adami:2020xkm}. In particular, three-dimensional gravity with torsion possesses a BTZ-like black hole solution \cite{Garcia:2003nm,Mielke:2003xx,Blagojevic:2003vn} whose thermodynamic properties have been analyzed in \cite{Blagojevic:2006jk,Blagojevic:2006hh,Blagojevic:2006nf} using different approaches. A gravity theory with both curvature and torsion can be formulated through the Mielke-Baekler (MB) gravity action \cite{Mielke:1991nn} which is described by the Einstein-Hilbert term, the cosmological constant term, the exotic Lagrangian \cite{Witten:1988hc} and a torsional term. Remarkably, for particular values of the MB parameters, the theory reproduces the teleparallel gravity.  As was shown in \cite{Blagojevic:2003uc}, the teleparallel theory has the same asymptotic structure as the Riemannian spacetime of General Relativity showing that the asymptotic structure seems not to depend on the underlying geometry, but only on the boundary conditions. Then, teleparallel gravity can be seen as an interesting toy model to explore the role of the torsion in the AdS/CFT correspondence \cite{Maldacena:1997re}. The extension to higher-spin and supersymmetry have then been explored in \cite{Peleteiro:2020ubv} and \cite{Salgado:2005pg,Giacomini:2006dr,Cvetkovic:2007sr}, respectively.

On the other hand, three-dimensional supergravity models \cite{Deser:1982sw,vanNieuwenhuizen:1985cx,Achucarro:1987vz,Achucarro:1989gm,Howe:1995zm,Banados:1996hi,Andringa:2009yc} are not only much simpler to handle but also useful to approach richer and higher-dimensional supergravities. In particular, supergravity without cosmological constant \cite{Marcus:1983hb} can be expressed as a Chern-Simons (CS) action invariant under the Poincaré superalgebra \cite{Achucarro:1987vz}. In presence of $\mathcal{N}=p+q$ supercharges, a well-defined Poincaré CS supergravity action requires to introduce automorphism generators which ensure the non-degeneracy of the bilinear invariant tensor \cite{Howe:1995zm,Caroca:2018obf}. Although three-dimensional supersymmetric gravity models with torsion have been explored in \cite{Giacomini:2006dr,Cvetkovic:2007sr}, a $\mathcal{N}$-extended supersymmetric CS formulation of the teleparallel gravity theory remains unexplored.

In this work, we present a teleparallel CS supergravity theory constructed from a novel superalgebra which can be seen as a supersymmetric extension of a deformed Poincaré algebra. The new symmetry has been denoted as teleparallel algebra since it allows us to construct a teleparallel gravity theory using the CS formalism. Although the teleparallel superalgebra is isomorphic to the $\mathfrak{osp}\left(2|1\right)\otimes\mathfrak{sp}\left(2\right)$ superalgebra, the supergravity theories based on them are quite different at the dynamics and geometric level. Indeed, the teleparallel supergravity theory presented here is characterized by a non-vanishing super-torsion in which the cosmological constant can be seen as a source for the torsion. Interestingly, the vanishing cosmological constant limit $\ell\rightarrow\infty$ leads us to the super Poincaré CS theory. The generalization of our results to $\mathcal{N}=p+q$ supersymmetries is also presented. Similarly to the AdS case \cite{Howe:1995zm}, the introduction of $\mathfrak{so}\left(p\right)\oplus\mathfrak{so}\left(q\right)$ generators are required in order to establish a well-defined flat limit.

The paper is organized as follows: in Section 2, we briefly discuss the teleparallel gravity and present its construction using the CS formulation and a teleparallel algebra. In section 3, we construct the minimal supersymmetric extension of the teleparallel CS gravity theory. Section 4 is devoted to the $\mathcal{N}=p+q$-extended generalization of our results. Section 5 concludes our work with discussions and comments about future developments.

\section{Three-dimensional teleparallel Chern-Simons gravity }\label{tel}

In this section, we present a brief review of the so-called teleparallel gravity in three spacetime dimensions. As it is well-known, this theory can be derived as a particular case of the MB gravity model \cite{Mielke:1991nn}, and is characterized by a non-vanishing torsion. The action for the MB gravity theory reads \cite{Mielke:1991nn,Cacciatori:2005wz}
\begin{equation}\label{MB}
    I_{\text{MB}}=aI_{1}+\Lambda I_{2}+\beta_{3} I_{3}+\beta_{4} I_{4}
\end{equation}
where $a,\Lambda,\beta_{3}$ and $\beta_{4}$ are constants and
\begin{eqnarray}
I_{1} & = & 2\int e_{a}R^{a}\,,\notag\\
I_{2} & = & -\frac{1}{3}\int \epsilon_{abc}e^{a}e^{b}e^{c}\,,\notag\\
I_{3} & = & \int \omega^a d\omega_a + \frac{1}{3}\epsilon^{abc}\omega_a \omega_b \omega_c\,,\notag\\
I_{4} & = &\int e_{a}T^{a}\,,\label{MBterms}
\end{eqnarray}
with 
\begin{eqnarray}
R^a&=&d\omega^a+\frac{1}{2}\epsilon^{abc}\omega_b\omega_c\,, \notag\\
T^a&=&de^{a}+\epsilon^{abc}\omega_b e_c\,, \label{curv}
\end{eqnarray}
being the Lorentz curvature and the torsion two-forms, respectively.

Three-dimensional teleparallel gravity \cite{Kawai:1993kvr,deAndrade:1997gka,Blagojevic:2003uc,Blagojevic:2003vn} can be obtained by fixing the parameters $(a, \Lambda, \beta_4)$ appearing in the MB gravity as
\begin{equation}\label{MBcts}
    a=\frac{1}{16\pi G}\,, \qquad \Lambda=-\frac{1}{4\pi G \ell^2}\,, \qquad \beta_{4}=-\frac{1}{8\pi G \ell}\,.
\end{equation}
Let us note that the parameter $\beta_{3}$ can be set to zero without lost of generality. Nevertheless, along this work we will keep it different from zero in order to maintain the exotic Lorentz term \cite{Witten:1988hc}. With this choice, the MB action \eqref{MB} takes the form
\begin{equation}
    I_{\text{TG}} = \frac{1}{16 \pi G}\int \tilde{\beta}_{3}\left(\omega^a d \omega_a + \frac{1}{3}\epsilon^{abc}\omega_a \omega_b \omega_c\right)+\left( 2 e_{a}R^{a}+\frac{4}{3 l^2}\epsilon_{abc}e^{a}e^{b}e^{c}-2e_{a}T^{a}\right)\,, \label{MBTel}
\end{equation}
where we have defined $\tilde{\beta}_{3}\equiv 16 \pi G \beta_{3}$.

In the following analysis, we will show that the teleparallel gravity action can alternatively be obtained as a CS gravity action invariant under a particular algebra, and whose variation leads to the equations of motion of the three-dimensional teleparallel gravity. Because of this property, we will refer to the mentioned symmetry as "teleparallel  algebra". This symmetry can be derived as a deformation of the Poincaré one and it is isomorphic to the $\mathfrak{so}(2,1)\otimes\mathfrak{so}(2,1)$ algebra.

The teleparallel algebra is spanned by the set of generators ($J_a$, $P_a$) which satisfy the following commutation relations:
\begin{eqnarray}
\left[J_a,J_b\right]&=&\epsilon_{abc} J^{c} \,, \notag \\
\left[J_a,P_b\right]&=&\epsilon_{abc} P^{c} \,, \notag \\
\left[P_a,P_b\right]&=&-\frac{2}{\ell}\epsilon_{abc} P^{c} \,, \label{algebra01} 
\end{eqnarray}
where $a,b=0,1,2$ are the Lorentz indices which are lowered and raised with the Minkowski metric $\eta_{ab}$ and $\epsilon_{abc}$ is the three-dimensional Levi-Civita tensor. On the other hand, the $\ell$ parameter is related to the cosmological constant through $\Lambda \propto - \frac{1}{\ell^2}$. In particular, the vanishing cosmological constant limit $\ell\rightarrow\infty$ applied to the teleparallel algebra reproduces the Poincaré algebra. Furthermore, let us note that the teleparallel algebra \eqref{algebra01}, under the following change of basis
\begin{eqnarray}
L_{a}&\equiv& J_{a} +\frac{\ell}{2} P_{a} \,,\qquad 
S_a\equiv -\frac{\ell}{2} P_{a}\,,\label{redefineads}
\end{eqnarray}
can be rewritten as two copies of the $\mathfrak{so}(2,1)$ algebra:
\begin{eqnarray}
\left[L_a,L_b\right]&=&\epsilon_{abc} L^{c} \,, \notag\\
\left[S_a,S_b\right] &=&\epsilon_{abc} S^{c} \,.
\end{eqnarray}

The general expression for a three-dimensional CS gravity action reads
\begin{equation}\label{CS-Action}
    I_{\text{CS}}[A]= \frac{k}{4 \pi} \int_{\mathcal{M}} \langle A d A+ \frac{2}{3} A^{3}  \rangle \,,
\end{equation}
where $A$ is the gauge connection one-form, $\langle \, , \, \rangle$ denotes the invariant trace and $k=\frac{1}{4G}$ is the CS level related to the gravitational constant $G$. For the sake of simplicity, here and in the sequel we will omit to write the wedge product. In particular, the gauge field connection one-form $A$ for the teleparallel algebra reads
\begin{equation}
    A= \omega^{a}J_{a}+e^{a} P_{a}\,,\label{one-form}
\end{equation}
where $\omega^{a}$ is the spin connection and $e^{a}$ is the dreibein. The corresponding curvature two-form $F=dA+\frac{1}{2}[A,A]$ is given by
\begin{equation}
    F= R^{a} J_{a}+\hat{T}^{a} P_{a}\,,\label{two-form}
\end{equation}
with
\begin{eqnarray}
 R^{a}&=&d\omega^{a}+\frac{1}{2}\epsilon^{abc}\omega_{b}\omega_{c} \,,\notag\\
 \hat{T}^{a}&=& T^{a}-\frac{1}{\ell}\epsilon^{abc} e_{b}e_{c} \,,\label{curvatures}
\end{eqnarray}
where $T^{a}$ is the usual torsion two-form defined in \eqref{curv}. Note that the vanishing cosmological constant limit $\ell\rightarrow\infty$ reproduces the Poincaré curvatures. On the other hand, the algebra \eqref{algebra01} admits a non-degenerate invariant bilinear form whose only non-vanishing components are given by
\begin{equation}
    \langle J_{a} J_{b} \rangle = \alpha_0 \,\eta_{a b}\,,\qquad \langle J_{a} P_{b} \rangle = \alpha_1 \,\eta_{a b}\,,\qquad \langle P_{a} P_{b} \rangle = -\frac{2\alpha_1}{\ell} \,\eta_{a b}\,.\label{invtensor}
\end{equation}
Here $\alpha_0$ and $\alpha_1$ are arbitrary constants which are related to the $\mathfrak{so}\left(2,1\right)$ constant through
$\alpha_0=\mu+\tilde{\mu}$ and $\alpha_1=-\left(2\tilde{\mu}\right)/{\ell}$. The non-degeneracy of the invariant tensor \eqref{invtensor} is preserved as long as  $\alpha_1\neq0$ and $2\alpha_0+\ell\alpha_1\neq0$. Such non-degeneracy is related to the requirement that the CS supergravity action involves a kinematical term for each gauge field.

A CS action invariant under the algebra \eqref{algebra01} can be written considering the non-vanishing components of the invariant tensor \eqref{invtensor} and the gauge potential one-form \eqref{one-form} in the general definition of the CS action \eqref{CS-Action},
\begin{equation}
    I_{\text{TG}} = \frac{1}{16 \pi G} \int_{\mathcal{M}} \bigg\{ \alpha_0 \left( \omega^a d \omega_a + \frac{1}{3}\epsilon^{abc}\omega_a \omega_b \omega_c
    \right) + \alpha_1 \left( 2 R_a e^a + \frac{4}{3\ell^{2}} \epsilon^{abc} e_a e_b e_c - \frac{2}{\ell}T^a e_a \right)\bigg\}\,,\label{TG}
\end{equation}
up to a boundary term.  The first term is the gravitational CS term with coupling constant $\alpha_0$ \cite{Witten:1988hc}. The second term proportional to the constant $\alpha_1$ contains the usual Einstein-Hilbert Lagrangian, a cosmological constant term and a torsional CS term. Comparing the previous action with \eqref{MBTel}, we realize that both actions are equal when the identification $\alpha_0=\tilde{\beta}_{3}$ and $\alpha_1=1$ is considered. One can see that the teleparallel CS action leads us to the Poincaré CS action in the vanishing cosmological constant limit $\ell\rightarrow\infty$. Due to the non-degeneracy of the invariant tensor, the corresponding equations of motion are given by:
\begin{eqnarray}
\delta e^{a} & : & \qquad0=\alpha_{1}\left(R_{a}-\frac{2}{\ell}\hat{T}_{a}\right),\nonumber \\
\delta\omega^{a} & : & \qquad0=\alpha_{0}R_{a}+\alpha_{1}\hat{T}_{a},\label{eom}
\end{eqnarray}
Since $\alpha_1\neq0$ and $\alpha_0\neq-\frac{\ell}{2}\alpha_1$, the above equations reduce to the vanishing of the curvature two-forms,
\begin{eqnarray}
R^{a} &=& 0\,, \notag\\
T^{a}-\frac{1}{\ell}\epsilon^{abc} e_{b}e_{c} &=& 0\,.\label{EOM}
\end{eqnarray}
Such equations of motion are geometrically dual to the AdS ones characterized by a Riemannian spacetime \cite{DeAndrade:2000sf}. Here, the CS gravity action \eqref{TG} describes a non-Riemannian geometry with a vanishing curvature and non-vanishing torsion $T^{a}\neq 0$. Thus, the CS action \eqref{TG} invariant under the algebra \eqref{algebra01} describes a gauge-invariant teleparallel gravity CS theory in three spacetime dimensions.


\section{On the minimal supersymmetric extension of teleparallel Chern-Simons gravity}\label{supertel}
In this section, we shall focus on a $\mathcal{N}=1$ supersymmetric extension of the teleparallel algebra in three spacetime dimensions.  The construction of a CS supergravity action based on this novel superalgebra is also presented. Interestingly, we will show that the CS teleparallel supergravity action is characterized by a non-vanishing super-torsion. 

\subsection{Teleparallel superalgebra}
A supersymmetric extension of the teleparellel algebra \eqref{algebra01} is spanned by a Lorentz generator $J_a$, a translational generator $P_a$ and a Majorana fermionic charge $Q_{\alpha}$. The super teleparallel generators satisfy the following non-vanishing (anti-)commutation relations:
\begin{eqnarray}
\left[J_a,J_b\right]&=&\epsilon_{abc} J^{c} \,, \notag \\
\left[J_a,P_b\right]&=&\epsilon_{abc} P^{c} \,, \notag \\
\left[P_a,P_b\right]&=&-\frac{2}{\ell}\epsilon_{abc} P^{c} \,, \notag \\
\left[J_a,Q_{\alpha}\right]&=&-\frac{1}{2}\left(\gamma_a\right)_{\alpha}^{\ \beta}Q_{\beta}\,, \notag \\
\left\{Q_{\alpha},Q_{\beta}\right\}&=&-\left(\gamma^aC\right)_{\alpha\beta}\left(\frac{2}{\ell}J_a+P_a\right)\,.\label{algebra02} 
\end{eqnarray}
Here $\alpha=1,2$ are spinorial indices, $\gamma^a$ are the Dirac matrices in three spacetime dimensions and $C$ is the charge conjugation matrix,
\begin{equation}
C_{\alpha\beta}=C^{\alpha\beta}=\begin{pmatrix}
0 & -1\\
1 & 0\\
\end{pmatrix}\, ,
\end{equation}
which satisfies $C\gamma^{A}=(C\gamma^{A})^{T}$ and $C^{T}=-C$. Let us note that the vanishing cosmological constant limit $\ell\rightarrow\infty$ leads us to the Poincaré superalgebra. On the other hand, the superalgebra \eqref{algebra02} can be written as the $\mathfrak{osp}\left(2|1\right)\otimes\mathfrak{sp}\left(2\right)$ superalgebra by considering the following identification of the generators:
\begin{eqnarray}
L_{a}&\equiv& J_{a} +\frac{\ell}{2} P_{a} \,,\qquad
S_a\equiv -\frac{\ell}{2} P_{a}\,,\qquad \mathcal{G}_{\alpha}\equiv \sqrt{\frac{\ell}{2}}Q_{\alpha}\,,\label{redefinesads}
\end{eqnarray}
where $\lbrace L_a,\mathcal{G}_{\alpha} \rbrace$ satisfy the $\mathfrak{osp}\left(2|1\right)$ superalgebra, while $S_a$ are $\mathfrak{sp}\left(2\right)$ generators,
\begin{eqnarray}
\left[L_a,L_b\right]&=&\epsilon_{abc} L^{c} \,, \notag\\
\left[S_a,S_b\right] &=&\epsilon_{abc} S^{c} \,, \notag \\
\left[L_a,\mathcal{G}_{\alpha}\right]&=&-\frac{1}{2}\left(\gamma_a\right)_{\alpha}^{\ \beta}\mathcal{G}_{\alpha} \,, \notag\\
\left\{\mathcal{G}_{\alpha},\mathcal{G}_{\beta}\right\}&=&-\left(\gamma^aC\right)_{\alpha\beta}L_a\,. 
\end{eqnarray}
Although the teleparallel algebra is isomorphic to the $\mathfrak{osp}\left(2|1\right)\otimes\mathfrak{sp}\left(2\right)$ superalgebra, the first one given by \eqref{algebra02} is quite different from the AdS superalgebra which, as we shall see, implies noticeable differences at the dynamics and geometric level. In particular, unlike the super AdS case, one can see that $\left[P,Q\right]=0$ and $\left[P,P\right]\sim P$. The latter implies, at the bosonic level, the presence of a non-vanishing torsion in which the cosmological constant can be seen as a source for the torsion.

\subsection{Chern-Simons supergravity action based on the teleparallel superalgebra}
Let $A=A^A T_A$ be the gauge connection one-form for the teleparellel superalgebra \eqref{algebra02},
\begin{equation}
    A= \omega^{a}J_{a}+e^{a} P_{a}+\bar{\psi}\,Q\,,\label{one-form2}
\end{equation}
where $\omega^a$ corresponds to the spin connection one-form, $e^a$ is the dreibein and $\psi$ is a Majorana fermionic one-form describing the gravitino. The curvature two-form reads
\begin{equation}
    F= \mathcal{R}^{a} J_{a}+\mathcal{T}^{a} P_{a}+\nabla\bar{\psi}\, Q\,,\label{two-form2}
\end{equation}
where
\begin{eqnarray}
 \mathcal{R}^{a}&=&R^{a}+\frac{1}{\ell}\bar{\psi}\gamma^{a}\psi \,,\notag\\
 \mathcal{T}^{a}&=& \hat{T}^{a}+\frac{1}{2}\bar{\psi}\gamma^{a}\psi \,,\notag\\
 \nabla\psi&=&d\psi+\frac{1}{2}\omega^{a}\gamma_a\psi\,.\label{curvatures2}
\end{eqnarray}
Here $\mathcal{R}^{a}$ describes the super-Lorentz curvature, $\mathcal{T}^{a}$ is a super-torsion and $\nabla\psi$ defines the covariant derivative of the gravitino. Furthermore, the bosonic curvatures 
$R^{a}$ and $\hat{T}^{a}$ were defined in \eqref{curvatures}. Let us note that the super Poincaré curvatures are recovered in the flat limit.

The teleparallel superalgebra \eqref{algebra02} admits the following non-degenerate invariant tensor,
\begin{eqnarray}
    \langle J_{a} J_{b} \rangle &=& \alpha_0 \,\eta_{a b}\,,\notag \\
    \langle J_{a} P_{b} \rangle &=& \alpha_1 \,\eta_{a b}\,,\notag \\
    \langle P_{a} P_{b} \rangle &=& -\frac{2\alpha_1}{\ell} \,\eta_{a b}\,, \notag \\
    \langle Q_{\alpha},Q_{\beta}\rangle &=& 2\left(\frac{2\alpha_0}{\ell}+\alpha_1\right)C_{\alpha\beta}\,, \label{invtensor2}
\end{eqnarray}
where $\alpha_0$ and $\alpha_1$ are arbitrary constants which are related to the $\mathfrak{osp}\left(2|1\right)\otimes\mathfrak{sp}\left(2\right)$ constants through
\begin{equation}
    \alpha_0=\mu+\tilde{\mu}\,, \qquad \alpha_1=-\frac{2\tilde{\mu}}{\ell}\,, \label{constants}
\end{equation}
with $\mu$ and $\tilde{\mu}$ being the coupling constants of the $\mathfrak{osp}\left(2|1\right)$ and $\mathfrak{sp}\left(2\right)$ algebras, respectively. In the flat limit we recover the non-vanishing components of the invariant tensor for the Poincaré superalgebra. In particular, there is no fermionic contributions to the exotic sector $\alpha_0$ \cite{Witten:1988hc} in the Poincaré limit.

Then, by considering the gauge connection one-form \eqref{one-form2} and the non-vanishing components of the invariant tensor \eqref{invtensor2} in the general expression of the CS action \eqref{CS-Action}, we find
\begin{eqnarray}
    I_{\text{TSG}} &=& \frac{1}{16 \pi G} \int_{\mathcal{M}} \bigg\{ \alpha_0 \left( \omega^a d \omega_a + \frac{1}{3}\epsilon^{abc}\omega_a \omega_b \omega_c -\frac{4}{\ell}\bar{\psi}\nabla\psi
    \right) \notag \\
    & &+ \alpha_1 \left( 2 R_a e^a + \frac{4}{3\ell^{2}} \epsilon^{abc} e_a e_b e_c - \frac{2}{\ell}T^a e_a -2\bar{\psi}\nabla\psi \right)\bigg\}\,.\label{TsG}
\end{eqnarray}
The CS action $I_{TSG}$ can be seen as a teleparallel supegravity action invariant under the teleparallel superalgebra \eqref{algebra02}. The CS supergravity action \eqref{TsG}, contains two independent sectors. The first term proportional to $\alpha_0$ describes a supersymmetric exotic action diverse to the one appearing in AdS supergravity \cite{Giacomini:2006dr}. In particular, unlike super AdS, the exotic term does not contain torsional term. Indeed, the torsion appears explicitly in the $\alpha_1$ sector along the Einstein-Hilbert term, the cosmological constant term and the fermionic kinetic term. Furthermore, the dreibein does not contribute to the covariant derivative of the fermionic gauge field. On the other hand, one can see that the vanishing cosmological constant limit $\ell\rightarrow\infty$ leads us to the Poincaré supergravity action whose exotic sector is no more supersymmetric. It is important to mention that the CS supergravity action \eqref{TsG} coincides with the most general supersymmetric gravity action presented in \cite{Giacomini:2006dr} for particular values of the parameters. 

Let us note that the corresponding field equations reads
\begin{eqnarray}
\delta e^{a} & : & \qquad0=\alpha_{1}\left(\mathcal{R}_{a}-\frac{2}{\ell}\mathcal{T}_{a}\right),\nonumber \\
\delta\omega^{a} & : & \qquad0=\alpha_{0}\mathcal{R}_{a}+\alpha_{1}\mathcal{T}_{a}, \nonumber \\
\delta\bar{\psi} & : & \qquad0=\frac{2\alpha_0}{\ell}\nabla\psi+\alpha_1\nabla\psi\label{eom2}
\end{eqnarray}
In particular, the non-degeneracy of the invariant tensor \eqref{invtensor2} requires $\alpha_1\neq0$ and $\alpha_0\neq-\frac{\ell}{2}\alpha_1$ which implies that the equations of motion are given by the vanishing of the curvature two-forms \eqref{curvatures2}. One can see that such supergravity theory corresponds to a supersymmetric extension of the teleparallel gravity and is characterized by a non-vanishing super-torsion,
\begin{equation}
    T^{a}+\frac{1}{2}\bar{\psi}\gamma^a\psi=\frac{1}{\ell}\epsilon^{abc}e_be_c\,.
\end{equation}
Naturally, in the flat limit $\ell\rightarrow\infty$ the super-torsion vanishes and we recover the super Poincaré field equations.

\section{\texorpdfstring{$\mathcal{N}$}{N}-extended teleparallel Chern-Simons supergravity theory}\label{Nsupertel}

In this section, we extend our construction to $\mathcal{N}=p+q$ supersymmetries. In particular, we show that the proper construction of an $\mathcal{N}$-extended teleparallel supergravity theory with a well-defined flat limit $\ell\rightarrow\infty$ requires the introduction of $\mathfrak{so}\left(p\right)\oplus\mathfrak{so}\left(q\right)$ automorphism generators as in $\left(p,q\right)$ AdS superalgebra \cite{Howe:1995zm}. Furthermore, the extra bosonic generators assures the non-degeneracy of the invariant tensor.

\subsection{\texorpdfstring{$\mathcal{N}$}{N}-extended teleparallel superalgebra}

A $\left(p,q\right)$ teleparallel superalgebra is spanned by a set of $p$ fermionic charges $Q_{\alpha}^{i}$, $i=1,\dots,p$,  and a complementary set of $q$ fermionic charges $Q_{\alpha}^{I}$, $I=1,\dots,q$, in addition to the bosonic generators $\lbrace J_a, P_a\rbrace$ and $p\left(p-1\right)/2+q\left(q-1\right)/2$ internal symmetry generators $Z^{ij}=-Z^{ji}$ and $Z^{IJ}=-Z^{JI}$. The $\left(p,q\right)$ teleparallel superalgebra satisfies the following non-vanishing (anti-)commutation relations
\begin{eqnarray}
\left[J_a,J_b\right]&=&\epsilon_{abc} J^{c} \,, \notag \\
\left[J_a,P_b\right]&=&\epsilon_{abc} P^{c} \,, \notag \\
\left[P_a,P_b\right]&=&-\frac{2}{\ell}\epsilon_{abc} P^{c} \,, \notag \\
\left[Z^{ij},Z^{kl}\right]&=&\delta^{jk}Z^{il}-\delta^{ik}Z^{jl}-\delta^{jl}Z^{ik}+\delta^{il}Z^{jk}\,,\notag\\
\left[Z^{IJ},Z^{KL}\right]&=&\delta^{JK}Z^{IL}-\delta^{IK}Z^{JL}-\delta^{JL}Z^{IK}+\delta^{IL}Z^{JK}\,,\notag\\
\left[J_a,Q_{\alpha}^{i}\right]&=&-\frac{1}{2}\left(\gamma_a\right)_{\alpha}^{\ \beta}Q_{\beta}^{i}\,, \notag \\
\left[J_a,Q_{\alpha}^{I}\right]&=&-\frac{1}{2}\left(\gamma_a\right)_{\alpha}^{\ \beta}Q_{\beta}^{I}\,, \notag \\
\left[P_a,Q_{\alpha}^{I}\right]&=&\frac{1}{\ell}\left(\gamma_a\right)_{\alpha}^{\ \beta}Q_{\beta}^{I}\,, \notag \\
\left[Z^{ij},Q_{\alpha}^{k}\right]&=&\delta^{jk}Q_{\alpha}^{i}-\delta^{ik}Q_{\alpha}^{j}\,, \notag \\
\left[Z^{IJ},Q_{\alpha}^{K}\right]&=&\delta^{JK}Q_{\alpha}^{I}-\delta^{IK}Q_{\alpha}^{J}\,, \notag \\
\left\{Q_{\alpha}^{i},Q_{\beta}^{j}\right\}&=&-\delta^{ij}\left(\gamma^aC\right)_{\alpha\beta}\left(P_a+\frac{2}{\ell}J_a\right)+\frac{2}{\ell}C_{\alpha\beta}Z^{ij}\,,\notag\\
\left\{Q_{\alpha}^{I},Q_{\beta}^{J}\right\}&=&-\delta^{IJ}\left(\gamma^aC\right)_{\alpha\beta}P_a-\frac{2}{\ell}C_{\alpha\beta}Z^{IJ}\,.\label{algebra03} 
\end{eqnarray}
Nevertheless, we require to introduce additional bosonic generators in order to recover, in the vanishing cosmological constant limit $\ell\rightarrow\infty$, the $\left(p,q\right)$ Poincaré algebra extended with the $\mathfrak{so}\left(p\right)\oplus\mathfrak{so}\left(q\right)$ automorphism algebra \cite{Howe:1995zm}. To this end, we extend the $\left(p,q\right)$ teleparallel superalgebra \eqref{algebra03} by $\mathfrak{so}\left(p\right)\oplus\mathfrak{so}\left(q\right)$ automorphism generators $S^{ij}=-S^{ji}$ and $S^{IJ}=-S^{JI}$ which satisfy
\begin{eqnarray}
\left[S^{ij},S^{kl}\right]&=&-\frac{2}{\ell}\left(\delta^{jk}S^{il}-\delta^{ik}S^{jl}-\delta^{jl}S^{ik}+\delta^{il}S^{jk}\right)\,,\notag\\
\left[S^{IJ},S^{KL}\right]&=&-\frac{2}{\ell}\left(\delta^{JK}S^{IL}-\delta^{IK}S^{JL}-\delta^{JL}S^{IK}+\delta^{IL}S^{JK}\right)\,. \label{autalg}
\end{eqnarray}
Then, we perform the following redefinition:
\begin{equation}
    T^{ij}=Z^{ij}-\frac{\ell}{2} S^{ij}\,,\qquad T^{IJ}=Z^{IJ}-\frac{\ell}{2} S^{IJ}\,,\label{redef}
\end{equation}
to eliminate $Z^{ij}$ and $Z^{IJ}$ and provide with a well-defined vanishing cosmological constant limit $\ell\rightarrow\infty$. With the redefinition \eqref{redef}, the direct sum of the $\left(p,q\right)$ teleparallel superalgebra and $\mathfrak{so}\left(p\right)\oplus\mathfrak{so}\left(q\right)$ automorphism algebra satisfies the teleparallel algebra \eqref{algebra01} along with \eqref{autalg} and
\begin{eqnarray}
\left[T^{ij},T^{kl}\right]&=&\delta^{jk}T^{il}-\delta^{ik}T^{jl}-\delta^{jl}T^{ik}+\delta^{il}T^{jk}\,,\notag\\
\left[T^{IJ},T^{KL}\right]&=&\delta^{JK}T^{IL}-\delta^{IK}T^{JL}-\delta^{JL}T^{IK}+\delta^{IL}T^{JK}\,,\notag\\
\left[T^{ij},S^{kl}\right]&=&\delta^{jk}S^{il}-\delta^{ik}S^{jl}-\delta^{jl}S^{ik}+\delta^{il}S^{jk}\,,\notag\\
\left[T^{IJ},S^{KL}\right]&=&\delta^{JK}S^{IL}-\delta^{IK}S^{JL}-\delta^{JL}S^{IK}+\delta^{IL}S^{JK}\,,\notag\\
\left[J_a,Q_{\alpha}^{i}\right]&=&-\frac{1}{2}\left(\gamma_a\right)_{\alpha}^{\ \beta}Q_{\beta}^{i}\,, \notag \\
\left[J_a,Q_{\alpha}^{I}\right]&=&-\frac{1}{2}\left(\gamma_a\right)_{\alpha}^{\ \beta}Q_{\beta}^{I}\,, \notag \\
\left[P_a,Q_{\alpha}^{I}\right]&=&\frac{1}{\ell}\left(\gamma_a\right)_{\alpha}^{\ \beta}Q_{\beta}^{I}\,, \notag \\
\left[T^{ij},Q_{\alpha}^{k}\right]&=&\delta^{jk}Q_{\alpha}^{i}-\delta^{ik}Q_{\alpha}^{j}\,, \notag \\
\left[T^{IJ},Q_{\alpha}^{K}\right]&=&\delta^{JK}Q_{\alpha}^{I}-\delta^{IK}Q_{\alpha}^{J}\,, \notag \\
\left\{Q_{\alpha}^{i},Q_{\beta}^{j}\right\}&=&-\delta^{ij}\left(\gamma^aC\right)_{\alpha\beta}\left(P_a+\frac{2}{\ell}J_a\right)+C_{\alpha\beta}\left(\frac{2}{\ell}T^{ij}+S^{ij}\right)\,,\notag\\
\left\{Q_{\alpha}^{I},Q_{\beta}^{J}\right\}&=&-\delta^{IJ}\left(\gamma^aC\right)_{\alpha\beta}P_a-C_{\alpha\beta}\left(\frac{2}{\ell}T^{IJ}+S^{IJ}\right)\,.\label{algebra04} 
\end{eqnarray}
The superalgebra given by \eqref{algebra01},\eqref{autalg} and \eqref{algebra04} shall be denoted as $\mathcal{N}$-extended teleparallel superalgebra and reproduces the $\left(p,q\right)$ Poincaré superalgebra extended with $\mathfrak{so}\left(p\right)\oplus\mathfrak{so}\left(q\right)$ automorphism algebra after considering the flat limit $\ell\rightarrow\infty$. The presence of automorphism generators in the Poincaré case are required in order to define non-degenerate invariant tensor \cite{Howe:1995zm}. As in the $\left(p,q\right)$ Poincaré superalgebra, the $S^{ij}$ and $S^{IJ}$ generators become central charges in the flat limit. However, although the $\mathcal{N}$-extended teleparellel superalgebra  presents a well-defined Poincaré limit, the (anti-)commutation relations are quite different from the AdS superalgebra. As we shall see, the $\mathcal{N}$-extended supergravity theory based on the $\mathcal{N}$-extended teleparallel superalgebra \eqref{algebra01},\eqref{autalg} and \eqref{algebra04} will imply rather different field equations.

Let us note that the $\mathcal{N}$-extended teleparallel superalgebra can be written as the direct sum of the $\mathfrak{osp}\left(2|p\right)\otimes\mathfrak{osp}\left(2|q\right)$ and the $\mathfrak{so}\left(p\right)\oplus\mathfrak{so}\left(q\right)$ automorphism algebra by considering the following identification of the generators:
\begin{eqnarray}
L_{a}&\equiv& J_{a} +\frac{\ell}{2} P_{a} \,,\qquad \quad \ \
S_a\equiv -\frac{\ell}{2} P_{a}\,,\qquad \,  \mathcal{G}_{\alpha}^{i}\equiv \sqrt{\frac{\ell}{2}}Q_{\alpha}^{i}\,,\notag\\
M^{ij}&\equiv&T^{ij}+\frac{\ell}{2}S^{ij}\,, \qquad \ \ B^{ij}\equiv -\frac{\ell}{2}S^{ij}\,, \qquad  \mathcal{G}_{\alpha}^{I}\equiv \sqrt{\frac{\ell}{2}}Q_{\alpha}^{I}\,, \notag\\
M^{IJ}&\equiv&T^{IJ}+\frac{\ell}{2}S^{IJ}\,, \qquad  B^{IJ}\equiv -\frac{\ell}{2}S^{IJ}\,,\label{redefinesads2}
\end{eqnarray}
where $\{ L_a,M^{ij},\mathcal{G}_{\alpha}^{i}\}$ and $\{ S_a,M^{IJ},\mathcal{G}_{\alpha}^{I}\}$ satisfy the $\mathfrak{osp}\left(2|p\right)$ and the $\mathfrak{osp}\left(2|q\right)$ superalgebra, respectively. On the other hand, $B^{ij}$ and $B^{IJ}$ are the respective $\mathfrak{so}\left(p\right)$ and $\mathfrak{so}\left(q\right)$ automorphism generators.

\subsection{Chern-Simons supergravity action and the \texorpdfstring{$\mathcal{N}$}{N}-extended teleparallel superalgebra}
Let us consider the gauge connection one-form $A$ for the $\mathcal{N}$-extended teleparellel superalgebra \eqref{algebra04},
\begin{equation}
    A= \omega^{a}J_{a}+e^{a} P_{a}+\frac{1}{2}A^{ij}T_{ij}+\frac{1}{2}A^{IJ}T_{IJ}+\frac{1}{2}C^{ij}S_{ij}+\frac{1}{2}C^{IJ}S_{IJ}+\bar{\psi}^{i}\,Q^{i}+\bar{\psi}^{I}\,Q^{I}\,,\label{one-form3}
\end{equation}
where the coefficients in front of the generators are the gauge field one-forms. The corresponding curvature two-form is given by
\begin{equation}
    F= \tilde{\mathcal{R}}^{a} J_{a}+\tilde{\mathcal{T}}^{a} P_{a}+\frac{1}{2}\tilde{F}^{ij}T_{ij}+\frac{1}{2}\tilde{F}^{IJ}T_{IJ}+\frac{1}{2}\tilde{G}^{ij}S_{ij}+\frac{1}{2}\tilde{G}^{IJ}S_{IJ}+\nabla\bar{\psi}^{i}\, Q^{i}+\nabla\bar{\psi}^{I}Q^{I}\,,\label{two-form3}
\end{equation}
where
\begin{eqnarray}
 \tilde{\mathcal{R}}^{a}&=&R^{a}+\frac{1}{\ell}\bar{\psi}^{i}\gamma^{a}\psi^{i} \,,\notag\\
 \tilde{\mathcal{T}}^{a}&=& \hat{T}^{a}+\frac{1}{2}\bar{\psi}^{i}\gamma^{a}\psi^{i}+\frac{1}{2}\bar{\psi}^{I}\gamma^{a}\psi^{I} \,,\notag\\
 \tilde{F}^{ij}&=& dA^{ij}+A^{ik}A^{kj}-\frac{2}{\ell}\bar{\psi}^{i}\psi^{j}\,,\notag\\
 \tilde{F}^{IJ}&=& dA^{IJ}+A^{IK}A^{KJ}+\frac{2}{\ell}\bar{\psi}^{I}\psi^{J}\,,\notag\\
 \tilde{G}^{ij}&=&dC^{ij}+A^{ik}C^{kj}+C^{ik}A^{kj}-\frac{2}{\ell}C^{ik}C^{kj}-\bar{\psi}^{i}\psi^{j} \,,\notag\\
 \tilde{G}^{IJ}&=&dC^{IJ}+A^{IK}C^{KJ}+C^{IK}A^{KJ}-\frac{2}{\ell}C^{IK}C^{KJ}+\bar{\psi}^{I}\psi^{J} \,,\notag\\
 \nabla\psi^{i}&=&d\psi^{i}+\frac{1}{2}\omega^{a}\gamma_a\psi^{i}+A^{ij}\psi^{j}\,,\notag\\
 \nabla\psi^{I}&=&d\psi^{I}+\frac{1}{2}\omega^{a}\gamma_a\psi^{I}-\frac{1}{\ell}e^{a}\gamma_a\psi^{I}+A^{IJ}\psi^{J} \,,\label{curvatures3}
\end{eqnarray}
and $R^{a}$, $\hat{T}^{a}$ are defined in \eqref{curvatures}. Analogously to the minimal case, the flat limit $\ell\rightarrow\infty$ reproduces the curvatures for the $\mathcal{N}$-extended Poincaré superalgebra.

One can show that the $\mathcal{N}$-extended teleparellel superalgebra \eqref{algebra04} admits the following non-vanishing components of the invariant tensor:
\begin{eqnarray}
    \langle J_{a} J_{b} \rangle &=& \alpha_0 \,\eta_{a b}\,,\notag \\
    \langle J_{a} P_{b} \rangle &=& \alpha_1 \,\eta_{a b}\,,\notag \\
    \langle P_{a} P_{b} \rangle &=& -\frac{2\alpha_1}{\ell} \,\eta_{a b}\,, \notag \\
    \langle T^{ij} T^{kl} \rangle &=& 2\alpha_0 \,\left(\delta^{il}\delta^{kj}-\delta^{ik}\delta^{jl}\right)\,, \notag \\
    \langle T^{IJ} T^{KL} \rangle &=& 2\alpha_0 \,\left(\delta^{IL}\delta^{KJ}-\delta^{IK}\delta^{JL}\right)\,, \notag \\
    \langle T^{ij} S^{kl} \rangle &=& 2\alpha_1 \,\left(\delta^{il}\delta^{kj}-\delta^{ik}\delta^{jl}\right)\,, \notag \\
     \langle T^{IJ} S^{KL} \rangle &=& -2\left(\frac{2\alpha_0}{\ell}+\alpha_1\right) \,\left(\delta^{IL}\delta^{KJ}-\delta^{IK}\delta^{JL}\right)\,, \notag \\
      \langle S^{ij} S^{kl} \rangle &=& -\frac{4\alpha_1}{\ell} \,\left(\delta^{il}\delta^{kj}-\delta^{ik}\delta^{jl}\right)\,, \notag \\
      \langle S^{IJ} S^{KL} \rangle &=& 2\left(\frac{4\alpha_0}{\ell^2}+\frac{2\alpha_1}{\ell}\right) \,\left(\delta^{IL}\delta^{KJ}-\delta^{IK}\delta^{JL}\right)\,, \notag \\
    \langle Q_{\alpha}^{i},Q_{\beta}^{j}\rangle &=& 2\left(\frac{2\alpha_0}{\ell}+\alpha_1\right)C_{\alpha\beta}\delta^{ij}\,,\notag \\
    \langle Q_{\alpha}^{I},Q_{\beta}^{J}\rangle &=& 2\alpha_1C_{\alpha\beta}\delta^{IJ}\,,\label{invtensor3}
\end{eqnarray}
where $\alpha_0$ and $\alpha_1$ are related to the $\mathfrak{osp}\left(2|p\right)\otimes\mathfrak{osp}\left(2|q\right)$ constants as in \eqref{constants} and to the $\mathfrak{so}\left(p\right)\oplus\mathfrak{so}\left(q\right)$ constants through
\begin{eqnarray}
\alpha_0=\rho+\tilde{\rho}\,,\qquad \qquad \alpha_1=-\frac{2\rho}{\ell}\,,
\end{eqnarray}
with $\rho$ and $\tilde{\rho}$ being the respective coupling constants of the $\mathfrak{so}\left(p\right)$ and $\mathfrak{so}\left(q\right)$ algebras. Let us note that the flat limit $\ell\rightarrow\infty$ leads us to the invariant tensor for the $\left(p,q\right)$ Poincaré superalgebra extended with $\mathfrak{so}\left(p\right)\oplus\mathfrak{so}\left(q\right)$ algebra \cite{Howe:1995zm}.

The CS action $I_{\text{TSG}}^{\mathcal{N}}$ for the $\mathcal{N}$-extended teleparellel superalgebra \eqref{algebra04} is obtained by considering the gauge connection one-form \eqref{one-form3} and the non-degenerate invariant tensor \eqref{invtensor3} in the general CS expression \eqref{CS-Action},
\begin{eqnarray}
    I_{\text{TSG}}^{\mathcal{N}} &=& \frac{1}{16 \pi G} \int_{\mathcal{M}} \bigg\{ \alpha_0 \left( \omega^a d \omega_a + \frac{1}{3}\epsilon^{abc}\omega_a \omega_b \omega_c+\mathcal{G}\left(A^{ij}\right)+\mathcal{G}\left(A^{IJ}\right) +\frac{4}{\ell}C^{IJ}\mathcal{F}^{IJ}+\frac{4}{\ell^2}\mathcal{G}\left(C^{IJ}\right)\right.\notag\\
    & & \left.-\frac{4}{\ell}\bar{\psi}^{i}\nabla\psi^{i}
    \right) + \alpha_1 \left( 2 R_a e^a + \frac{4}{3\ell^{2}} \epsilon^{abc} e_a e_b e_c - \frac{2}{\ell}T^a e_a -2C^{ij}\mathcal{F}^{ij}+2C^{IJ}\mathcal{F}^{IJ}-\frac{2}{\ell}\mathcal{G}\left(C^{ij}\right) \right.\notag\\
   & &\left. +\frac{2}{\ell}\mathcal{G}\left(C^{IJ}\right)-2\bar{\psi}^{i}\nabla\psi^{i}-2\bar{\psi}^{I}\nabla\psi^{I}\right)\bigg\}\,.\label{NTsG}
\end{eqnarray}
where
\begin{eqnarray}
    \mathcal{G}\left(A^{ij}\right)&=&A^{ij}dA^{ji}+\frac{2}{3}A^{ik}A^{km}A^{mi}\,,\notag\\
    \mathcal{G}\left(A^{IJ}\right)&=&A^{IJ}dA^{JI}+\frac{2}{3}A^{IK}A^{KM}A^{MI}\,,\notag\\
     \mathcal{G}\left(C^{ij}\right)&=&C^{ij}dC^{ji}-\frac{4}{3\ell}C^{ik}C^{km}C^{mi}\,,\notag\\
    \mathcal{G}\left(C^{IJ}\right)&=&C^{IJ}dC^{JI}-\frac{4}{3\ell}C^{IK}C^{KM}C^{MI}\,,\notag\\
    \mathcal{F}^{ij}&=&dA^{ij}+A^{ik}A^{kj}-\frac{1}{\ell}C^{ik}A^{kj}-\frac{1}{\ell}A^{ik}C^{kj}\,,\notag\\
    \mathcal{F}^{IJ}&=&dA^{IJ}+A^{IK}A^{KJ}-\frac{1}{\ell}C^{IK}A^{KJ}-\frac{1}{\ell}A^{IK}C^{KJ}\,.
\end{eqnarray}
The $\mathcal{N}$-extended teleparallel CS supergravity action \eqref{NTsG} contains two independent sectors proportional to $\alpha_0$ and $\alpha_1$. The term proportional to $\alpha_0$ contains the exotic Lagrangian plus contribution of the gravitini and internal symmetry gauge fields. On the other hand, the term proportional to $\alpha_1$ contains the teleparallel gravity terms present in \eqref{TG} plus contribution of the automorphism gauge fields and gravitini. In the vanishing cosmological constant limit $\ell\rightarrow\infty$ the CS action reproduces the $\left(p,q\right)$ Poincaré supergravity extended with $SO\left(p\right)\times SO\left(q\right)$ automorphism gauge fields \cite{Howe:1995zm}. In such limit, the gravitini and automorphism gauge fields do not contribute anymore to the exotic sector.

The equation of motions derived from the CS supergravity action \eqref{NTsG} are given by
\begin{eqnarray}
\delta e^{a} & : & \qquad0=\alpha_{1}\left(\tilde{\mathcal{R}}_{a}-\frac{2}{\ell}\tilde{\mathcal{T}}_{a}\right)\,,\nonumber \\
\delta\omega^{a} & : & \qquad0=\alpha_{0}\tilde{\mathcal{R}}_{a}+\alpha_{1}\tilde{\mathcal{T}}_{a}\,, \nonumber \\
\delta\bar{\psi}^{i} & : & \qquad0=\frac{2\alpha_0}{\ell}\nabla\psi^{i}+\alpha_1\nabla\psi^{i}\,, \nonumber \\
\delta\bar{\psi}^{I} & : & \qquad0=\alpha_1\nabla\psi^{I}\,, \nonumber \\
\delta A^{ij} & : & \qquad0=\alpha_0\tilde{F}^{ij}+\alpha_1 \tilde{G}^{ij}  \,, \nonumber \\
\delta A^{IJ} & : & \qquad0=\alpha_0\left(\tilde{F}^{IJ}-\frac{2}{\ell}\tilde{G}^{IJ}\right)-\alpha_1 \tilde{G}^{IJ} \,, \nonumber \\
\delta C^{ij} & : & \qquad0=\alpha_1\left( \tilde{F}^{ij}-\frac{2}{\ell}\tilde{G}^{ij}\right)  \,, \nonumber \\
\delta C^{IJ} & : & \qquad0=\frac{2\alpha_0}{\ell}\left(\tilde{F}^{IJ}-\frac{2}{\ell}\tilde{G}^{IJ}\right)+\alpha_1 \left(\tilde{F}^{IJ}-\frac{2}{\ell}\tilde{G}^{IJ}\right) \,.
\label{eom3}
\end{eqnarray}
Let us note that for $\mathcal{N}=\left(1,1\right)$, the second equation reproduces the field equations for the supersymmetric extension of gravity with torsion \cite{Cvetkovic:2007sr}.  On the other hand, the non-degeneracy of the invariant tensor \eqref{invtensor3}, which requires $\alpha_1\neq0$ and $\alpha_0\neq-\frac{\ell}{2}\alpha_1$, implies that the equations of motion reduce to the vanishing of the curvature two-forms \eqref{curvatures3}. As in the minimal case, the $\mathcal{N}$-extended teleparallel supergravity theory is characterized by a non-vanishing super-torsion,
\begin{eqnarray}
T^{a}+\frac{1}{2}\bar{\psi}^{i}\gamma^{a}\psi^{i}+\frac{1}{2}\bar{\psi}^{I}\gamma^{a}\psi^{I}=\frac{1}{\ell}\epsilon^{abc} e_{b}e_{c}\,.
\end{eqnarray}

\section{Conclusions}

In this work we have presented a teleparallel supergravity theory in three spacetime dimensions considering the CS formalism. To this end we have first shown that a teleparallel CS gravity action can be constructed using the gauge connection one-form for a particular deformation of the Poincaré algebra, which we have denoted as teleparallel algebra. The supersymmetric extension of the teleparallel algebra has then been considered to obtain a teleparallel supergravity action characterized by a non-vanishing super-torsion. In presence of $\mathcal{N}=p+q$ supersymmetry charges, the consistent construction of a teleparallel supergravity action with a well-defined flat limit requires to consider the direct sum of the $\left(p,q\right)$ teleparallel superalgebra and the $\mathfrak{so}\left(p\right)\oplus\mathfrak{so}\left(q\right)$ automorphism algebra. The latter ensures having a non-degenerate invariant tensor which is related to the physical requirement that the CS action involves a kinematical term for each field. Remarkably, both teleparallel and AdS description of (super)gravity reproduces the Poincaré (super)gravity theory in the vanishing cosmological limit. However, in the teleparallel formulation of (super)gravity, the flat limit is responsible of the vanishing (super)-torsion.

The results presented here could serve as a starting point for diverse further studies. In particular, the $\mathcal{N}=2$ super teleparallel gravity theory could be useful to elucidate a non-relativistic counterpart of the present theory. Non-relativistic supergravity theories have just been explored this last decade with a growing interest \cite{Andringa:2013mma,Bergshoeff:2015ija,Bergshoeff:2016lwr,Ozdemir:2019orp,deAzcarraga:2019mdn,Ozdemir:2019tby,Concha:2019mxx,Concha:2020tqx,Concha:2020eam}. In particular a teleparallel version of the extended Newton-Hooke supergravity \cite{Ozdemir:2019tby} is unknown and could bring valuable information about the role of the torsion in a non-relativistic environment and its relation to Newtonian supergravity (work in progress).

On the other hand, the CS formulation of (super)gravity is useful to study asymptotic symmetry and obtain a canonical realization of infinite-dimensional symmetry. It would be then interesting to study appropriate boundary conditions to our teleparallel (super)gravity theory and analyze its boundary dynamics. One could expect to recover the same asymptotic structure than the one obtained in the supersymmetric extension of gravity with torsion \cite{Cvetkovic:2007sr}.

Another aspect that deserves further investigation is the Maxwellian version of the teleparallel supergravity. A Maxwell generalization of three-dimensional gravity with torsion has been presented in \cite{Adami:2020xkm}. Such construction has been obtained from a deformation of the so-called Maxwell algebra \cite{Schrader:1972zd,Bacry:1970ye} which has proved to have several applications in the gravity context \cite{Concha:2013uhq,Concha:2014vka,Salgado:2014jka,Hoseinzadeh:2014bla,Caroca:2017izc,Concha:2018zeb,Aviles:2018jzw,Concha:2020sjt,Chernyavsky:2020fqs,Concha:2020ebl}. A supersymmetric version of the deformed Maxwell algebra could be considered to construct a Maxwellian teleparallel supergravity theory in three spacetime dimensions. At the bosonic level, the study of the black hole solution and thermodynamics of the Maxwellian teleparallel gravity could bring valuable information about the physical implications of a non-vanishing torsion in Maxwell gravity theory.

\section*{Acknowledgment}

This work was funded by the National Agency for Research and Development ANID (ex-CONICYT) - PAI grant No. 77190078 (P.C.). This work was supported by the Research project Code DIREG$\_$09/2020 (R.C. and P.C.) of the Universidad Católica de la Santisima Concepción, Chile. R.C. and P.C. would like to thank to the Dirección de Investigación and Vice-rectoría de Investigación of the Universidad Católica de la Santísima Concepción, Chile, for their constant support.


\bibliographystyle{fullsort.bst}
 
\bibliography{Teleparallel_supergravity}

\end{document}